# The Fallacy of Borda Count Method - Why it is Useless with Group Intelligence and Shouldn't be Used with Big Data including Banking Customer Services

Hao Wang*

CEO & Founder, Ratidar Technologies LLC, Beijing, China

**Abstract.** Borda Count Method is an important theory in the field of voting theory. The basic idea and implementation methodology behind the approach is simple and straight forward. Borda Count Method has been used in sports award evaluations and many other scenarios, and therefore is an important aspect of our society. An often ignored ground truth is that online cultural rating platforms such as Douban.com and Goodreads.com often adopt integer rating values for large scale public audience, and therefore leading to Poisson/Pareto behavior. In this paper, we rely on the theory developed by Wang from 2021 to 2023 to demonstrate that online cultural rating platform rating data often evolve into Poisson/Pareto behavior, and individualistic voting preferences are predictable without any data input, so Borda Count Method (or, Range Voting Method) has intrinsic fallacy and should not be used as a voting theory method.

## 1. INTRODUCTION

Borda Count Method is an important voting theory research subject. The idea of Borda Count Method is to give n votes for the n-star candidate, and evaluate the final score in sum total. For example, online cultural rating platforms adopt range voting methodology for their UGC content, which can be deemed as a variant of the Borda Count Method. This voting methodology leads to the following discovery: On Douban.com and Goodreads.com, scientists [1][2][3][4] discovered that the 5-star cultural products (books / movies / music) usually receive 4 times more raters than 1 star cultural products. This is actually compatible with intuition because the more popular a movie is, the more audience it will enlist.

Borda Count Method is reported to be adopted in various customer services evaluation processes as well. For example, in the banking system, some customer service systems ask the customers to rate the banking accountants using 1 to n stars. The Golden Globe Award evaluation procedure also uses Borda Count Method.

In 2021, Wang [1] discovered that without using data inputs, we could predict individualistic user preference on online cultural rating platforms by assuming the rating behavior follows Poisson / Pareto behavior. The reason why rating behavior follows Poisson / Pareto is mostly by experimental observation on large-scale social science datasets such as MovieLens [6] and LDOS-CoMoDa [7] datasets.

The implications of the series of zeroshot learning methods proposed by Wang [1][2][3][4] are that Borda Count Method as a voting mechanism leads to results that are predictable without any reference to the actual voting data with a fair level of accuracy.

In this paper, we discuss this discovery in details and disclose the deep insight we've found as a consequence. We suggest policy makers to eliminate the Borda Count Method / Range Voting Method as a voting mechanism from any voting evaluation process.

## 2. RELATED WORK

Borda Count Method is an important voting theory topic. The method was proposed in 1770 by French scientist and has bee circulating in the next centuries, although it is believed the first adoption of the method is much earlier.

An often neglected technical area by social scientists is recommender sytem. Recommender system have experienced an outburst of innovation and invention, including shallow models [8][9][10] and deep neural networks [11][12][13]. Another issue has arised that attracts the attention from the research circle is fairness [14][15].

In 2023, Wang analyzed the evolution behavior of online rating platform data [16] and discovered that the rating behavior follows non-homogeneous Poisson process. Prior to that, ZeroMat [1], DotMat [2], PoissonMat [3], Pareto Pairwise Ranking [4], among a series of algorithms were proposed to generate data-agnostic results for recommendation and personalization tasks.

---

* Corresponding author: haow85@live.com

## 3. Data Agnostic Models

Although not explicitly elaborated in his research works, Wang [1][2][3][4] discovered that the rating values of cultural products mostly follow Poisson / Pareto behavior: The 5-star movie on Douban.com usually enlists 4 times more audience than 1-star movie. The 5-star is the de facto best movie rated by this voting mechanism.

Based on this observation, ZeroMat [1] was invented. The algorithm is developed in the following way:

$$P(R|U, V, \sigma_U, \sigma_V) \sim \prod_{i=1}^{N} \prod_{j=1}^{M} (U_i^T \cdot V_j) N(0, \sigma_U) N(0, \sigma_V)$$

Applying Stochastic Gradient Descent Algorithm to solve the loss function, we acquire the following update rules by setting the standard deviation of the prior distribution of the user feature and item feature vectors (modeled as normal distributions and standard deviation fixed to be 1):

$$U_i = U_i + \varphi \times \left( \frac{V_j}{U_i^T \cdot V_j} - 2 \times U_i \right)$$

$$V_j = V_j + \varphi \times \left( \frac{V_j}{V_j^T \cdot U_i} - 2 \times V_j \right)$$

Unbelievably, the algorithm leads to unimaginably good results when compared with heuristics and other algorithmic models with full datasets.

Another notable invention is Pareto Pairwise Ranking [5]. The algorithm jumps out of the realm of matrix factorization framework, and extends the idea of zeroshot learning to the field of learning to rank. The algorithm is formulated below :

$$L = \sum_{i=1}^{n} \sum_{j=1}^{m} \sum_{k=1}^{m} P(R_{i,j} > R_{i,k}) I(R_{i,j} > R_{i,k})$$

By Zipf Law phenomenon, namely the n-star movie receives n-1 times more audience than 1-star movie (a naturally formed Poisson / Pareto style evolutionary mechanism), we have the following formula :

$$L = \sum_{i=1}^{n} \sum_{j=1}^{m} \sum_{k=1}^{m} \frac{1}{(U_i^T \cdot V_j - U_i^T \cdot V_k)^\alpha} I(R_{i,j} > R_{i,k})$$

Applying Stochastic Gradient Decent algorithm to solve the loss function, we have the following update rules :

$$U_i = U_i + \varphi \alpha \frac{V_j - V_k}{U_i^T \cdot V_j - U_i^T \cdot V_k}$$

$$V_j = V_j + \varphi \alpha \frac{U_i}{U_i^T \cdot V_j - U_i^T \cdot V_k}$$

$$V_k = V_k - \varphi \alpha \frac{U_i}{U_i^T \cdot V_j - U_i^T \cdot V_k}$$

The accuracy result of Pareto Pairwise Ranking is also remarkably good, with a very high level of fairness when evaluated for popularity bias.

Fig.1 [5] shows the accuracy result of Pareto Pairwise Ranking v.s ZeroMat and other methods. Fig.2 [5] demonstrates the fairness metrics computed in the experiment. The dataset that the scientist has been using is MovieLens dataset. The author also tested his algorithms on the LDOS-CoMoDa dataset, the details of which are omitted in this publication.

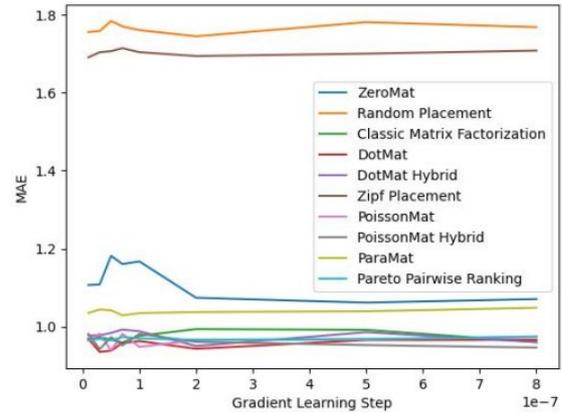

**Fig.1** MAE comparison among 10 algorithms on MovieLens dataset

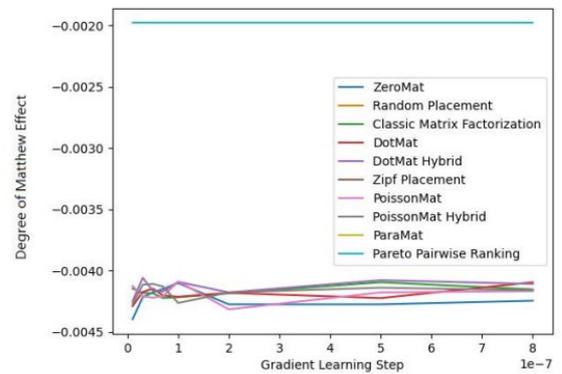

**Fig.2** Fairness comparison among 10 algorithms on MovieLens dataset

We notice that among the 10 algorithms, classic matrix factorization algorithm - an algorithm with full data input loses its competitive edge when compared against a host of data agnostic algorithms

The latest cold-start learning algorithm invented by Wang [17] is Skellam Rank. The algorithm builds a data-agnostic model using Poisson process and Skellam distribution. The loss function of the algorithm is formulated as below :

$$L = \sum_{i=1}^{n} \sum_{w=1}^{n} \sum_{j=1}^{m} \sum_{k=1}^{m} P(R_{i,j} > R_{w,k}) I(R_{i,j} > R_{i,k})$$

Using Poisson process, Skellam distribution and Zipf law to expand the loss function. We have a more precise and clear definition of our loss function :

$$L = \sum_{i=1}^{n} \sum_{w=1}^{n} \sum_{j=1}^{m} \sum_{k=1}^{m} e^{-(E_i+E_w)} \left( \frac{E_i}{E_w} \right)^{\frac{E_i - E_w}{2}} I_k(2\sqrt{E_i E_w})$$

The function of E is defined as follows :

$$E_i = \frac{1}{n}\sum_{j=1}^{n}\frac{R_{i,j}}{n}, E_w = \frac{1}{n}\sum_{j=1}^{n}\frac{R_{w,j}}{n}$$

, while the function of I is defined as follows :

$$I_k = \sum_{t=0}^{\infty}\frac{(-1)^t}{t!\,(t+E_i-E_w)!}\left(\sqrt{E_iE_w}\right)^{2t+E_i-E_w}$$

Applying Stochastic Gradient Descent to the loss function, we obtain the following pseudo-code :

```
Function Skellam-Rank :
1    Read input data into the user item rating matrix R
2    For iter in 1: max_iter_number:
3        User_sample = sample users from R
4        γ = constant
5        For user i in user_sample:
6            U = random sample from uniform distribution
7            Item_sample = sample items from user i's item rating list
8            Item_list = sorted item samples in decreasing rating values
9            V = random sample from uniform distribution
10           For j in 1: max_index_Item_list -1:
11               For k in j+1: max_index_Item_list:
12                   If R[i, j] > R[i, k]:
13                       U_i = U_i - γ ∂L/∂U_i
14                       U_w = U_w - γ ∂L/∂U_w
15                       V_j = V_j - γ ∂L/∂V_j
16                       V_k = V_k - γ ∂L/∂V_w
16           Reconstruct R by dot products of U and V
```

In Wang [17], the author compared Skellam Rank with 9 other algorithms, including algorithm with full data input, and data-agnostic algorithms without any input data. Fig.3 - Fig.4 demonstrate the results on MovieLens dataset, and Fig.5 - Fig.6 demonstrate the results on LDOS-CoMoDa dataset.

From the figures, we notice that Skellam Rank is a strong competitor with other algorithms, especially when the enumeration parameter is becoming larger. Without much endeavour, it would be very easy to identify that the classic matrix factorization is an out-dated and easily defeated technology, when compared with data agnostic models.

In general, zeroshot learning algorithms are highly competitive with classic matrix factorization algorithm of full data input, which means the invention (with Probabilistic Matrix Factorization as a later reformulation) is basically a failed attempt of earlier years to model recommendation and personalization behaviors.

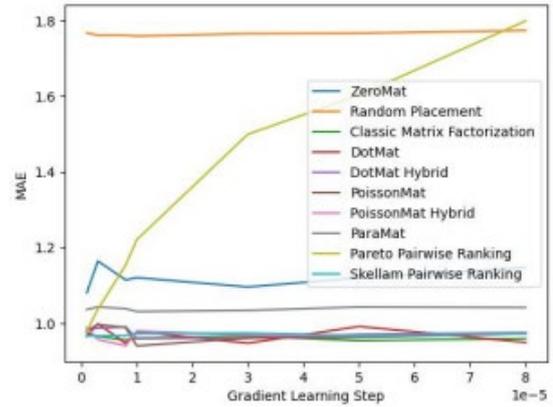

**Fig. 3** MAE comparison among 10 algorithms on MovieLens dataset

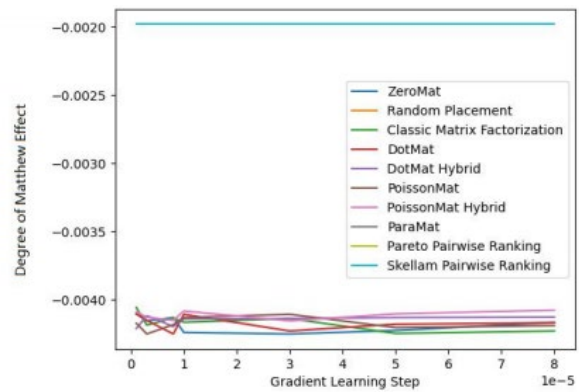

**Fig. 4** Fairness comparison among 10 algorithms on MovieLens dataset

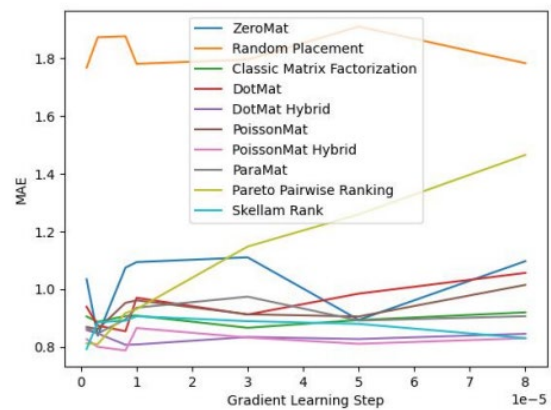

**Fig. 5** MAE comparison among 10 algorithms on LDOS-CoMoDa dataset

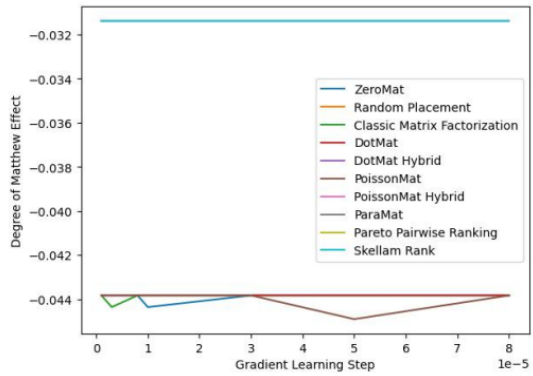

**Fig.6** Fairness comparison among 10 algorithms on LDOS-CoMoDa dataset

## 4. Borda Count Method

This might not be the first debut of the method, but Jean-Charles de Borda proposed the idea of Borda Count Method in the year of 1770. The idea is for audience to give the most favorite candidate n tickets, the second most favorite candidate n-1 tickets, etc. This idea has been promulgated in the field of customer services disguised in the form of range voting in various economic sectors such as banking systems.

As we've overviewed our inventions for zeroshot learning , we notice that the result of Borda Count Method can be predicted without reference to the actual voting data. To elaborate, raters on web platforms such as Douban.com and Goodreads.com give 5 stars to top cultural products, and 4 stars to the second place, etc. We could classify cultural products of same rating value (as accumulated calculated by weighted sum) as a single candidate and reformulate the problem as a Borda Count Method problem, since the n-star value products usually receive n-1 times more raters than 1-star value products. We notice with time flying by, the final rating preference of each individual can be estimated fairly accurately without reference to voting data, as detailed in the previous section.

This phenomenon invalidates the voting theory of Borda Count Method in the big data setting - we've successfully shown that the voting results of Borda Count Method can be approximated fairly accurately without even referring to the voting data. Our proof is not a stand-alone discovery, but 8 publications with 1 *Best Oral Presentation Award* at AIBT 2023 (Wang [17]).

The implications of this discovery is that important customer services system such as the one with banking system is intrinsically flawed, and they should be uninstalled to reflect the true opinions of the customers. Otherwise, the opinions of the customers would not be reflected in the system because the rating results could be estimated without reference to any actual data.

## 5. Conclusion

In this paper, we overviewed 8 algorithmic inventions proposed by the author and discovered that the Borda Count Method and its variant Range Voting Method are invalid voting models for big data including banking customer services. We justified our model with mathematical formulations and experimental data, and proved our conclusion.

In future work, we'd like to test the validity of other voting theories, and possibly, come up with new theories and technologies, so we would be able to create a new school of thought that better our society.